\documentclass[
    ,final            % use final for the camera ready runs
  ]
  {aipproc}

\layoutstyle{6x9}

\begin{document}

\title{GRBlog: A Database for Gamma-Ray Bursts}

\author{Robert Quimby}{
  address={McDonald Observatory, University of Texas at Austin, Austin, TX 78712}
}

\author{Erin McMahon}{
  address={McDonald Observatory, University of Texas at Austin, Austin, TX 78712}
}

\author{Jeremy Murphy}{
  address={McDonald Observatory, University of Texas at Austin, Austin, TX 78712}
}

\begin{abstract}
GRBlog is an on-line database providing researchers with quick access to all information reported in the GCN Circulars. Users of the GRBlog web site (grad40.as.utexas.edu/grblog.php) can search the circulars and produce afterglow light curve plots, or compile data tables. The site also offers advanced search capabilities to aide in statistical studies or comparative research. Most of the GCNs have already been entered into the GRBlog database, with the remainder to follow shortly.
\end{abstract}

\maketitle

%%%%%%%%%%%%%%%%%%%%%%%%%%%%%%%%%%%%%%%%%%%%
%% MAINMATTER
%%%%%%%%%%%%%%%%%%%%%%%%%%%%%%%%%%%%%%%%%%%%

\section{Motivation}
The GRB Coordinates Network (GCN), created by Scott Barthelmy\cite{barthelmy:gcn}, has proven to be an invaluable tool for the rapid dissemination of gamma-ray burst observations, and vital to the success of afterglow studies. The Notices portion of the GCN conveys burst locations and types automatically, in a format easily parsed at robotic observatories. The details of any ensuing observations are then sent out to the GRB community via the GCN Circulars. Although the Circulars are automatically distributed, the free format nature of these messages can make it difficult for the community to sort through all the information and find the data they need to address their research questions.

The first important step to organize the GCN Circulars was made by Jochen Greiner, who created a web site with the Circulars sorted by bursts and maintains a master table listing key observational data for each burst\cite{greiner:web}. We have taken the further step of inserting all the data from the GCN Circulars into a relational database with a web interface called GRBlog\footnote{See http://grad40.as.utexas.edu/grblog.php}. In this format, the data can easily be retrieved by the entire community. Users of GRBlog can freely download observational data compiled from a number of Circulars in a tabular format, or produce plots of afterglow light curves.

\section{The Database}
We have implemented GRBlog using a PostgreSQL database, with a PHP web interface. To insert a new Circular into the database, we first download the ASCII file from the GCN archives. We then mark up that text with a custom formatting language, similar to Latex. For example, we would convert the passage ``\texttt{R = 18.5 mag}'' to ``\texttt{$\backslash$filter\{R\} = $\backslash$mag\{18.5\} mag}''. The marked up message is then passed through a PHP interpreter which converts it to HTML, and saves the marked up data to various tables (e.g. a table for optical observations). There are some common patterns as to how one might express a given observation in the GCN Circulars. We have identified several of these and have in place some routines to automatically mark up the Circulars. We then manually check the messages and make adjustments as necessary.

Once the messages have been marked up and their data appropriately stored in the database, PHP routines sort through the data and create HTML web pages on the fly.

\subsection{Message View}
The Message View is used to display the actual text of a Circular. The GRBlog default home page for example displays the 20 most recent messages. When more than one message is selected, only the first few lines of the text are displayed with a link to the full message. Each message begins with a title bar giving both the Circular subject line, and a link to the original Circular in the GCN archive. Data collected during the markup process is used to automatically classify the message, and a message category icon (e.g. eye glasses for an optical observation report) is shown on the right. Some of the marked up data will be converted to a link to another page. For example, the burst name GRB030329 will serve as a link to the 030329 burst page.

When a single message is selected, it will begin with a formated version of the Circular header followed by the list of authors (each name is a link to a list of messages the author is credited with). The message text will follow with embedded links to bursts,  message references, and websites. Any tables will be presented as an in-line HTML table. At the end of the message text, the number of citations to the current message will be shown with a link to a list of those messages. Below the citation count are tables summarizing any observational data presented in the message.

\subsection{Burst View}
The Burst View is used to quickly convey all the available observational data for a given burst, or list of bursts. Shown in a table for each burst are the fluence, RA and DEC, number of optical observations, etc., as available. For HETE bursts, the trigger number will be determined on the fly by looking for messages reporting both the HETE trigger number and the GRB name, and a link will be provided to the appropriate HETE Burst Page\cite{HETE:bursts}. When afterglow observations are available, detection of the transient is given as ``yes'' or ``no'' which serves as a link to the full light curve (it will be ``yes'' only if a message in the GRBlog database gives a magnitude for the OT which is not an upper limit).

\begin{figure}
  \includegraphics[height=.4\textheight]{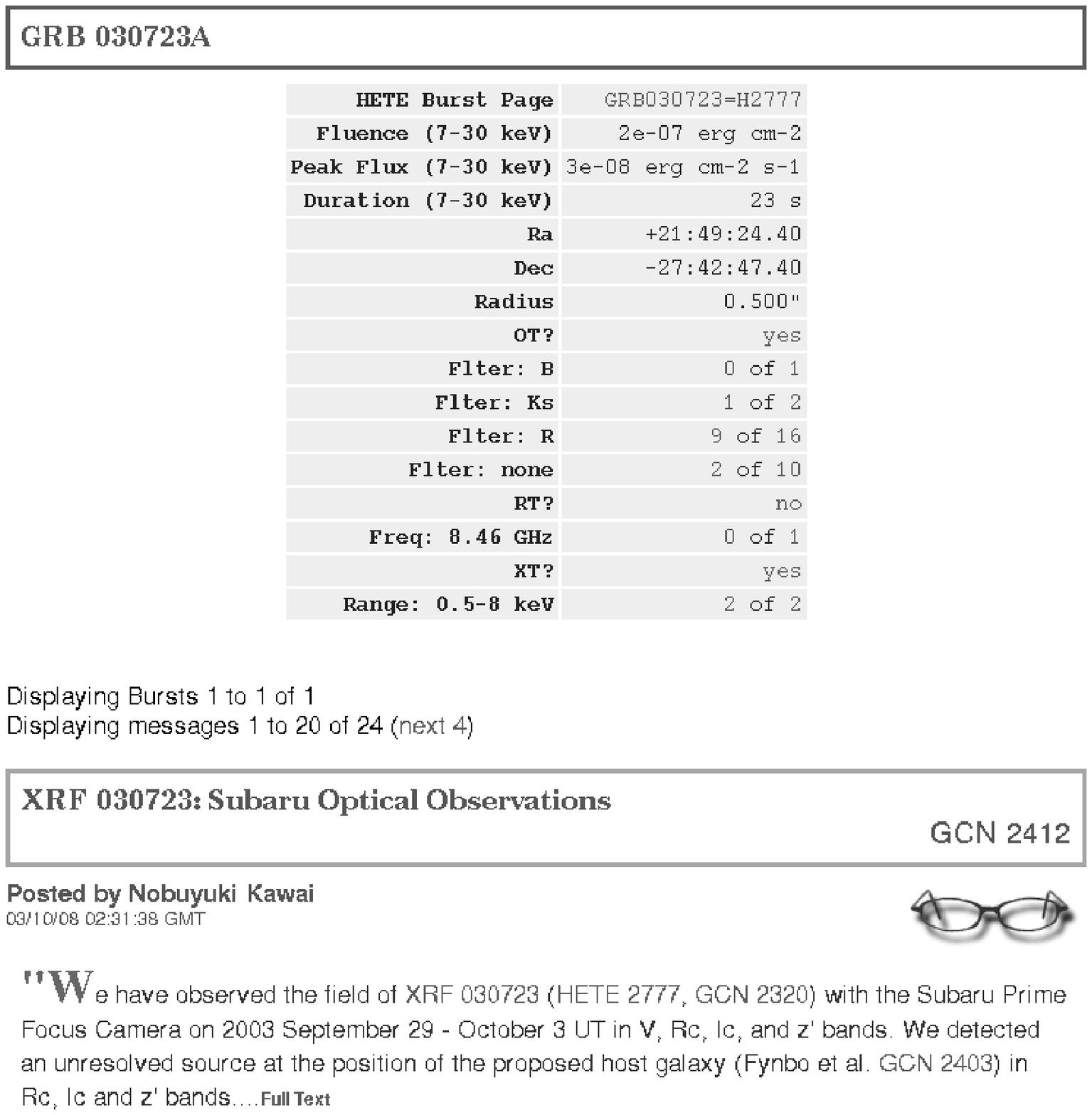}\label{burst_page}
  \caption{Detail from a screen shot of the GRB030723 burst page. Values in the table serve as links to other pages and light curve plots. Below the burst data summary are the introductions to the GCN circulars related to this burst (only one is shown). View the full page at http://grad40.as.utexas.edu/grblog.php?view=burst\&GRB=030723.}
\end{figure}

When a single burst is selected, all of the messages related to that burst are displayed below the data summary table. An example of the burst table is shown in Fig. \ref{burst_page}.

\subsection{Data View}
\begin{figure}
  \includegraphics[height=.34\textheight]{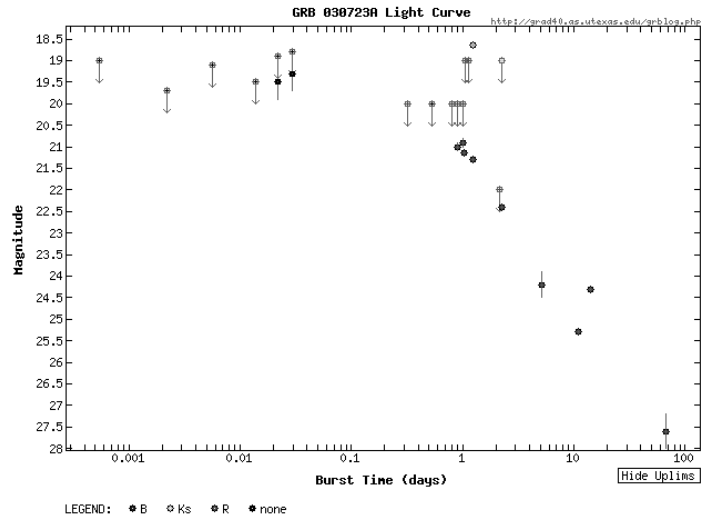}\label{light_curve}
  \caption{Lightcurve of GRB030723 generated automatically by GRBlog.}
\end{figure}

Data can be extracted from a collection of messages and displayed either in a table, or as a plot. The source for each data point is given in the tables, and can be viewed from plots by clicking on a given data point. As an example, to view the optical light curve for GRB030723, first go to the burst page, then click on ``yes'' in the table next to ``OT?'' (Fig. \ref{burst_page}). The data will be plotted using the magnitudes and observation times given in the Circulars (see Fig. \ref{light_curve}); they will not be converted to the same reference system. To see what reference system a given observation employs, look for the ``refsys'' column in the data table. Light curves can also be made from radio or X-ray observations. GRB data for collections of bursts can also be displayed using the search form. 

\section{Searching}
The web site includes a search form to allow users to find messages, bursts, or data which meet specific criteria. One can search for all messages that contain the word ``ROTSE'' for example, or to find all bursts with afterglows discovered within one hour of the burst. The search form has fields to limit the search to specific burst names, author names, fluence ranges, durations, optical magnitudes, spectra, etc., and all of the fields can be combined to produce complex searches.

\section{Future}
We will soon have all of the GCN Circulars entered into our database, and we will continue to add new Circulars as they become available. Once the database is complete with respect to the GCNs, we will consider expanding the database to include IAU Circulars, and possibly papers from astro-ph.

Although difficult to implement, using a markup language to convert prose into a machine readable format has proven quite successful. We note here that in the Swift era when the volume of GCN Circulars increases significantly, it may be important to agree on a common format for presenting observations to the community. If the GCNs reports were marked up to clearly identify data in a manner similar to that used by GRBlog, then authors would continue to have great freedom in how they express their results, but the results would be easier for the community to parse. 

%%%%%%%%%%%%%%%%%%%%%%%%%%%%%%%%%%%%%%%%%%%%%%%%
%% BACKMATTER
%%%%%%%%%%%%%%%%%%%%%%%%%%%%%%%%%%%%%%%%%%%%%%%%

\begin{theacknowledgments}
We thank Bradley E. Schaefer, J. Craig Wheeler, Pawan Kumar, Peter H\"oflich, Chris Gerardy, and Alin Panaitescu for valuable suggestions on the layout and functionality of the GRBlog website. This work is supported in part by NSF AST 0098644.
\end{theacknowledgments}

%% If the bibliography is produced without BibTeX comment out the

\bibliographystyle{aipproc}   % if natbib is available

\bibliography{refs}

\IfFileExists{\jobname.bbl}{}
 {\typeout{}
  \typeout{******************************************}
  \typeout{** Please run "bibtex \jobname" to optain}
  \typeout{** the bibliography and then re-run LaTeX}
  \typeout{** twice to fix the references!}
  \typeout{******************************************}
  \typeout{}
 }

\end{document}